\documentclass[%
 aip,
 amsmath,amssymb,
 reprint,%
]{revtex4-1}

\usepackage{graphicx}
\usepackage{dcolumn}
\usepackage{bm}

\usepackage[utf8]{inputenc}
\usepackage[T1]{fontenc}
\usepackage{mathptmx}
\usepackage{latexsym}
\usepackage{amssymb}
\usepackage{amsthm}
\usepackage{enumitem}
\usepackage{amsmath}
\usepackage{color}
\usepackage{soul}
\newcommand{\angstrom}{\text{\normalfont\AA}}

\soulregister\cite7
\soulregister\citep7
\soulregister\onlinecite7
\soulregister\ref7

\begin{document}

\preprint{AIP/123-QED}

\title[Time- and momentum-resolved photoemission studies using time-of-flight momentum microscopy at a free-electron laser]{Time- and momentum-resolved photoemission studies using \\time-of-flight momentum microscopy at a free-electron laser}



\author{D. Kutnyakhov}
 \email{dmytro.kutnyakhov@desy.de.}
 \homepage{https://uniexp.desy.de/}
\affiliation{Deutsches Elektronen-Synchrotron DESY, 22607 Hamburg, Germany}
\author{R. P. Xian}
\author{M. Dendzik}
\affiliation{Fritz Haber Institute of the Max Planck Society, 14195 Berlin, Germany}
\author{M. Heber}
\affiliation{Deutsches Elektronen-Synchrotron DESY, 22607 Hamburg, Germany}
\author{F. Pressacco}
\affiliation{Physics Department and Centre for Free-Electron Laser Science (CFEL), University of Hamburg, 22761 Hamburg, Germany}
\affiliation{The Hamburg Centre for Ultrafast Imaging (CUI), 22761 Hamburg, Germany}
\author{S. Y. Agustsson}
\affiliation{Institut f\"ur Physik, Johannes Gutenberg-Universit\"at Mainz, 55128 Mainz, Germany}
\author{L. Wenthaus}
\affiliation{Deutsches Elektronen-Synchrotron DESY, 22607 Hamburg, Germany}
\author{H. Meyer}
\author{S. Gieschen}
\author{G. Mercurio}
\author{A. Benz}
\affiliation{Physics Department and Centre for Free-Electron Laser Science (CFEL), University of Hamburg, 22761 Hamburg, Germany}
\author{K. B\"uhlman}
\author{S. D\"aster}
\author{R. Gort}
\affiliation{Laboratorium f\"ur Festk\"orperphysik, ETH Z\"urich, 8093 Z\"urich, Switzerland}
\author{D. Curcio}
\author{K. Volckaert}
\author{M. Bianchi}
\affiliation{Department of Physics and Astronomy, Interdisciplinary Nanoscience Center (iNANO), Aarhus University, 8000 Aarhus C, Denmark}
\author{Ch. Sanders}
\affiliation{Central Laser Facility, STFC Rutherford Appleton Laboratory, Harwell OX11 0QX, United Kingdom}
\author{J. A. Miwa}
\author{S. Ulstrup}
\affiliation{Department of Physics and Astronomy, Interdisciplinary Nanoscience Center (iNANO), Aarhus University, 8000 Aarhus C, Denmark}
\author{A. Oelsner}
\affiliation{Surface Concept GmbH, 55124 Mainz, Germany}
\author{C. Tusche}
\author{Y.-J. Chen}
\affiliation{Forschungszentrum J\"ulich GmbH, Peter Gr\"unberg Institut (PGI-6), 52428 J\"ulich, Germany}
\affiliation{Fakult\"at f\"ur Physik, Universit\"at Duisburg-Essen, 47057 Duisburg, Germany}
\author{D. Vasilyev}
\author{K. Medjanik}
\affiliation{Institut f\"ur Physik, Johannes Gutenberg-Universit\"at Mainz, 55128 Mainz, Germany}
\author{G. Brenner}
\author{S. Dziarzhytski}
\author{H. Redlin}
\author{B. Manschwetus}
\affiliation{Deutsches Elektronen-Synchrotron DESY, 22607 Hamburg, Germany}
\author{S. Dong}
\author{J. Hauer}
\author{L. Rettig}
\affiliation{Fritz Haber Institute of the Max Planck Society, 14195 Berlin, Germany}
\author{F. Diekmann}
\affiliation{Institut f\"ur Experimentelle und Angewandte Physik, Christian-Albrechts-Universit\"at zu Kiel, 24098 Kiel, Germany}
\author{K. Rossnagel}
\affiliation{Deutsches Elektronen-Synchrotron DESY, 22607 Hamburg, Germany}
\affiliation{Institut f\"ur Experimentelle und Angewandte Physik, Christian-Albrechts-Universit\"at zu Kiel, 24098 Kiel, Germany}
\affiliation{Ruprecht-Haensel-Labor, Christian-Albrechts-Universit\"at zu Kiel and Deutsches Elektronen-Synchrotron DESY, 24098 Kiel and 22607 Hamburg, Germany}
\author{J. Demsar}
\author{H.-J. Elmers}
\affiliation{Institut f\"ur Physik, Johannes Gutenberg-Universit\"at Mainz, 55128 Mainz, Germany}
\author{Ph. Hofmann}
\affiliation{Department of Physics and Astronomy, Interdisciplinary Nanoscience Center (iNANO), Aarhus University, 8000 Aarhus C, Denmark}
\author{R. Ernstorfer}
\affiliation{Fritz Haber Institute of the Max Planck Society, 14195 Berlin, Germany}
\author{G. Sch\"onhense}
\affiliation{Institut f\"ur Physik, Johannes Gutenberg-Universit\"at Mainz, 55128 Mainz, Germany}
\author{Y. Acremann}
\affiliation{Laboratorium f\"ur Festk\"orperphysik, ETH Z\"urich, 8093 Z\"urich, Switzerland}
\author{W. Wurth}
\affiliation{Deutsches Elektronen-Synchrotron DESY, 22607 Hamburg, Germany}
\affiliation{Physics Department and Centre for Free-Electron Laser Science (CFEL), University of Hamburg, 22761 Hamburg, Germany}

\date{\today}

\begin{abstract}
Time-resolved photoemission with ultrafast pump and probe pulses is an emerging technique with wide application potential. Real-time recording of non-equilibrium electronic processes, transient states in chemical reactions or the interplay of electronic and structural dynamics offers fascinating opportunities for future research. Combining valence-band and core-level spectroscopy with photoelectron diffraction for electronic, chemical and structural analysis requires few 10~fs soft X-ray pulses with some 10~meV spectral resolution, which are currently available at high repetition rate free-electron lasers. The PG2 beamline at FLASH (DESY, Hamburg) provides a high pulse rate of 5000~pulses/s, 60~fs pulse duration and 40~meV bandwidth in an energy range of 25--830~eV with a photon beam size down to 50~microns in diameter. We have constructed and optimized a versatile setup commissioned at FLASH/PG2 that combines FEL capabilities together with a multidimensional recording scheme for photoemission studies. We use a full-field imaging momentum microscope with time-of-flight energy recording as the detector for mapping of 3D band structures in ($k_x$, $k_y$, $E$) parameter space with unprecedented efficiency. Our instrument can image full surface Brillouin zones with up to 7~\angstrom$^{-1}$ diameter in a binding-energy range of several~eV, resolving about $2.5\times10^5$ data voxels. As an example, we present results for the ultrafast excited state dynamics in the model van der Waals semiconductor WSe$_2$.
\end{abstract}

\maketitle

X-ray free-electron lasers (XFELs) are a remarkable development in the range of tools for scientific experimentation. They are characterized by much enhanced peak brightness, many orders of magnitude in comparison to other X-ray sources, pulse durations on the order of a few tens of femtoseconds, the possibility for polarization control and wavelength tuning in a broad energy range, and multi-colour operation mode. Consequently, they have created tremendous new opportunities for experimental studies. In particular, high repetition rate extreme-ultraviolet (XUV) and soft X-ray FELs such as FLASH at DESY (Hamburg)\cite{Ackermann2007,Tiedtke2009} offer unique possibilities for time-resolved photoemission spectroscopy (trPES) to study the ultrafast electron dynamics in condensed matter systems like low-dimensional materials, molecular crystals or quantum materials. For example, the solid state physics community has demonstrated increasing interest in quantum materials in the last few years due to their intriguing emergent properties\citep{Gedik2017}. In particular, the area of topological states of matter is a very hot topic with different classes of topological insulators\citep{Hsieh2009,Hasan2010} in two and three dimensions as well as with 3D Weyl semimetals\citep{Marsi2018,Caputo2018}, displaying bulk band gaps, bulk Dirac or Weyl points, topological surface Dirac cones or Fermi arcs, and their associated characteristic spin textures, respectively. Coexistence of topological phases with the intriguing properties of other collective phases like Kondo lattices, high-T$_c$ superconductivity, charge- and spin-density waves, and Mott-Hubbard insulating states is another very interesting subject\citep{Dzero2010,Miyamachi2017,Lee2006,Lee2013}. The novel experimental approach presented here expands momentum-resolved photoemission spectroscopy (PES) using momentum microscopy with time-of-flight (ToF) energy recording \cite{Schoenhense2015,Medjanik2017} into the time domain as time-resolved momentum microscopy (trMM)\citep{Agustsson2019,Pressacco2019,Scholz2019}. Recently, parallel spin detection \citep{Kutnyakhov2016,Elmers2016,Schoenhense2017}, circular (or linear) dichroism in the angular distribution CDAD (LDAD)\citep{Chernov2015,Fedchenko2019}, X-ray photoelectron spectroscopy (XPS) \citep{Dendzik2019}, and X-ray photoelectron diffraction (XPD) \citep{Curcio2019,Schoenhense2018} have also been successfully performed with this instrument. The combination of these measurement techniques in a single experiment can directly probe and disentangle the fundamental interactions underlying different emergent properties of complex quantum materials. Adding temporal information on the sub-picosecond timescale opens a new path to direct determination of the couplings between the electronic, spin and lattice degrees of freedom. Such studies are of central importance to quantum materials, since the couplings are not only relevant to non-equilibrium dynamics, but also determine the materials' ground-state properties. 

While sources based on laser-driven high harmonic generation (HHG) are well-suited for trMM studies of condensed matter samples in the vacuum ultraviolet (VUV) spectral range with excellent time resolution\citep{Buss2019,Ploetzing2016,Puppin2019,Eich2014,Corder2018,Mills2019,Rohde2018,Haarlammert2009,Mathias2010}, high repetition rate FELs are capable of providing ultrashort pulses in a much broader spectral range from the XUV to the soft and even tender X-ray regime with convenient spectral tunability. Higher photon energies for trMM enable studies covering an extended range in momentum space. The tunability of the photon energy of the probe beam allows access to the full 3D momentum information and provides the means to study surface vs. bulk phenomena, because the inelastic mean free path of the photoelectrons and hence the probing depth can be tuned by changing the kinetic energy of the photoelectrons. Furthermore, the broad energy range of the FEL also allows to probe many core level states so that valence band measurements can be complemented by time-resolved XPS (trXPS). XPS is a hugely powerful technique that has been able to shed light onto many different types of samples and questions\citep{Huefner2003}. Its potential for time-resolved studies is highly promising\citep{Hellmann2012a,Siefermann2014}, but yet to be fully exploited. Moreover, the k-resolution in trMM implies that not only angle-integrated trXPS is measured, but that this is done in a k- and thus angle-resolved manner, yielding trXPD patterns with the possibility to study ultrafast and element-specific changes in the geometric structure along with the corresponding changes in the electronic structure.

A fundamental problem in trPES studies on targets with high electron density is the appearance of space-charge effects, if too many photoelectrons are emitted within one ultrashort photon pulse. In order to avoid these effects, the number of emitted photoelectrons has to be kept below a certain limit. As a consequence, it is necessary to use the most efficient detection schemes in combination with high repetition rate sources. Owing to its 3D data acquisition scheme, the ToF momentum microscope is well-suited to carry out time- and momentum-resolved PES studies on (semi-)conducting solid samples. 

The aim of this paper is to demonstrate the performance of our instrument on the model case of carrier dynamics in WSe$_2$. We have studied bulk WSe$_2$ with 2H stacking (2H-WSe$_2$), which belongs to the class of van der Waals semiconducting transition metal dichalcogenides (scTMDCs)\citep{Wilson1969,Grasso1986}. The family of scTMDCs compounds recently regained scientific interest due to their superior optoelectronic properties and the possibilities for their convenient integration with current nanotechnology\citep{Wang2012,Wachter2017}, whereas their rich underlying light-matter interactions and multifaceted quasiparticle dynamics demand in-depth investigations\citep{Wang2018}. We report the ultrafast dynamics of the excited states in the conduction band as well as the core levels following resonant optical excitation to the \textit{A} exciton transition in bulk WSe$_2$ crystals\citep{Frindt1963}. Furthermore, we observe excited state populations in the K (and K$'$) valleys, which have previously been reported to be localized on individual WSe$_2$ layers\citep{Bertoni2016}. The unique combination of high time- and momentum-resolution and the broad photon-energy tuning range of the momentum microscope setup offers unique possibilities for a comprehensive study of the carrier dynamics by rapid volumetric band-mapping experiments without the need for angular scanning\citep{Wang2018,Traving1997,Finteis1999,Riley2014}. The present paper describes the details of the experimental setup and procedures used for the first time- and momentum-resolved photoemission measurements at FLASH, DESY (Hamburg).

\begin{figure*}
\center{\includegraphics[width=17cm]{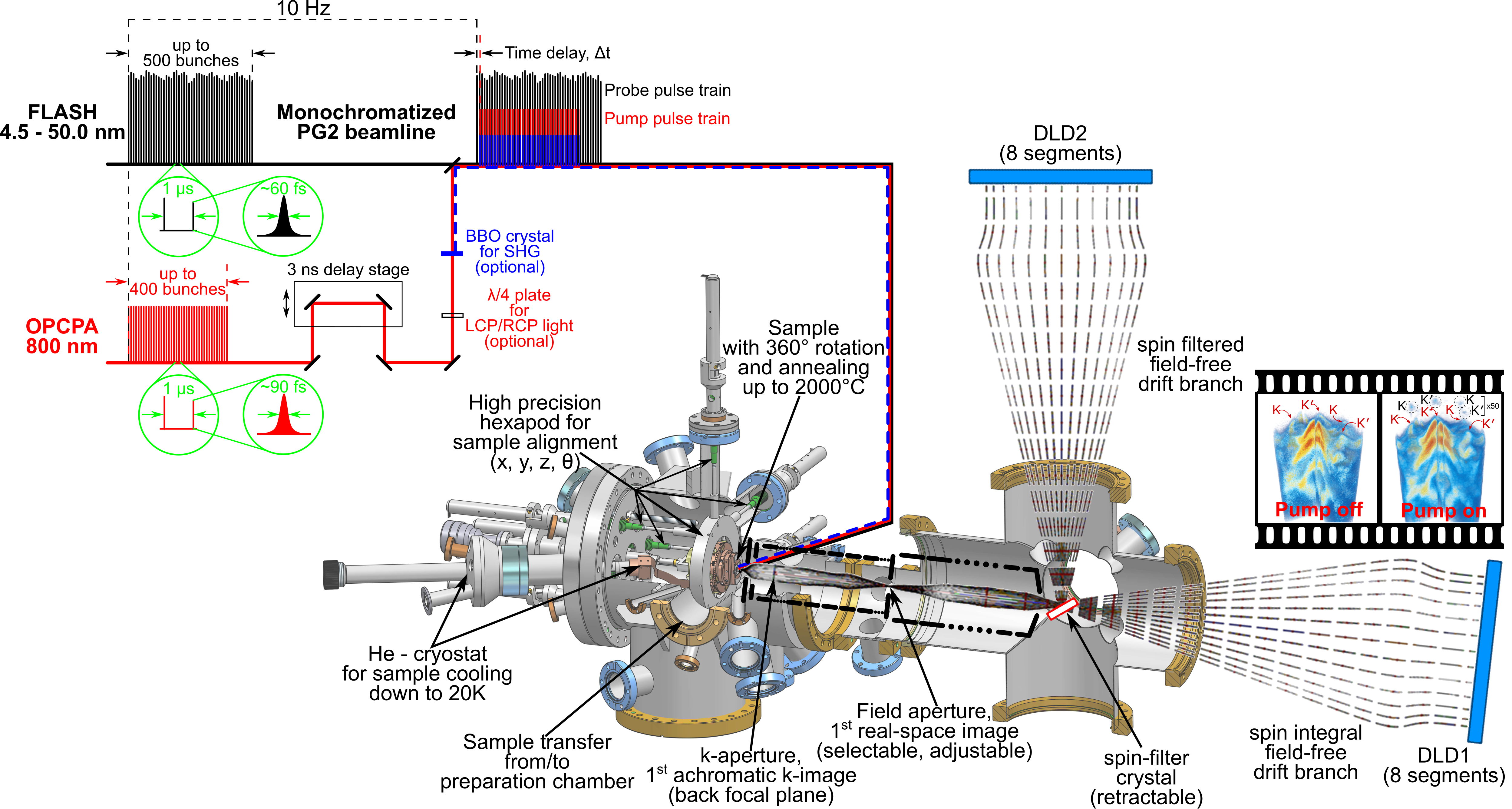}}
\caption{(color online) Simplified overview of the FLASH pulse structure, the PG2 beamline with synchronized pump laser (top left) and the experimental setup (middle) used to perform time-resolved momentum microscopy and acquire volumetric band-mapping movies (right inset) by means of a multidimensional recording scheme as described in detail in the text.
\label{Set-Up}
 }
\end{figure*}

\section{\label{sec:Experiment}Experimental method}
\subsection{\label{sec:FEL+IR} FEL and IR beams}

Near-infrared (NIR) pump--soft-X-ray probe experiments have been performed at the PG2 beamline \cite{Martins2006,Wellhoefer2007,Gerasimova2011} of FLASH, DESY (Hamburg). The schematic of the facility and the experimental setup is shown in Fig.\,\ref{Set-Up}. FLASH produces ultrashort XUV and soft X-ray pulses in a wavelength range of 4--50~nm (fundamental) and down to approximately 1.5~nm (third harmonic), which corresponds to a total photon energy range of 25--830~eV. 
We present here photoemission data measured at 36.5~eV (fundamental) for valence band studies and 109.5~eV (third harmonic) for core level studies. The number of FLASH pulses was set to 500 pulses per macrobunch with 1~MHz intra-bunch repetition rate and 10~Hz macrobunch rate, resulting in effectively 5000 probe pulses per second with 33~$\mu $J (0.33~$\mu $J) pulse energy in the fundamental (third harmonic). The radiation was monochromatized by a 200 lines/mm plane grating monochromator with a fix-focus constant (c$_{\text{ff}}$) of 1.6.
Such a dispersion causes a tilt of the wavefront and thus elongation of the 60~fs intrinsic FEL pulse duration. 
In order to preserve reasonable time resolution of the experiment, we closed vertical baffles before the beamline monochromator to decrease the number of illuminated grooves of the grating. This results in $\sim$180~fs ($\sim$130~fs) FWHM pulse duration along the beam propagation direction for trMM (trXPS) measurements, respectively. The exit slit was set to 400~$\mu $m (100~$\mu $m) providing 32~meV (40~meV)  spectral width for valence band (core level) measurements, respectively. The pointing stability of the self-amplified spontaneous emission-type (SASE) FEL beam was improved by using two remote apertures of 1 mm in diameter after the last undulator in the tunnel. This also reduces the FEL pulse energy down to 1--2~$\mu $J for the fundamental. The pulse energy was attenuated further by: i) a gas attenuator (filled with 1.5--3$\times10^{-2}$~mbar of N$_2$), ii) solid filters (Al foils with thickness of 2060~nm and 960~nm for valence band studies and Zr with 920~nm and ZrB$_2$ with 200~nm thickness for core level studies) and iii) the beamline transmission down to 3~fJ pulse energy for valence band studies and  to 100~fJ in the case of core level studies. Finally, about $\sim$500 ($\sim$6000)  photons per pulse arrived on the sample for valence band (core level) measurements, respectively. The footprint of the soft X-ray beam on the sample surface was measured to be about 300~$\mu $m $\times$ 150~$\mu $m (see Fig.\,\ref{Chessy}(b)) using the PEEM mode of the momentum microscope (see  Sec.\,\ref{sec:Overlap} below).

The pump laser system based on an optical parametric chirped pulse amplification (OPCPA) delivered $s$- or $p$-polarized (user defined) laser pulses with the central wavelength of 800~nm, of up to $\sim$15~$\mu$J pulse energy and $\sim$90~fs FWHM pulse duration\citep{Redlin2011}. A combination of neutral density filters and motorized half wave-plate followed by a polarizer could be used to tune the excitation density over several orders of magnitude, which allows to cover the range from the one-photon photoemission limit\cite{Scholz2019} to high fluence experiments in the range of more than 5~mJ/cm$^2$ such as optically induced phase transitions\cite{Pressacco2019}, or other driven collective phenomena. The laser pulse train was fully synchronized with the soft X-ray pulses~\cite{Schulz2015} and matched the complex macrobunch time pattern of the FEL. Each macrobunch from the laser system contained up to 400 pulses. Hence, by using 500 FEL pulses within the pulse train, the experiment provides $\sim$100 "unpumped" soft X-ray pulses after each optical pump pulse train. The photoemission data recorded with the unpumped probe pulses were used for reference and background correction. 
The temporal synchronization of the FEL and the optical laser pulses is additionally controlled by a streak camera, which monitors the arrival time of the optical laser pulses and of the synchrotron radiation generated by the electron bunches, when they are deflected via a dipole magnet into the beam dump. The two beams are combined using a holey mirror and propagate collinearly to the sample at an angle of $22^{\circ}$ with respect to the sample surface. The measured footprint of the optical pump laser on the sample is $\sim$ 510~$\mu $m $\times$ 115~$\mu$m, as shown in Fig.\,\ref{Chessy}(c) (using the PEEM mode of the momentum microscope described below in Sec.\,\ref{sec:Overlap}).  In order to probe spin-, valley- and layer-polarized excited states of WSe$_2$ we used both linearly and circularly polarized pump pulses, with an incident fluence of 6.1~mJ/cm$^2$ (total absorbed fluence of about 1.7~mJ/cm$^2$). Control over the pump pulse polarization was achieved by a motorized quarter-wave plate for switching between $s$, $p$, $\sigma^-$, and $\sigma^+$ polarizations (see Fig.\,\ref{Set-Up}). The pump pulse photon energy was tuned to 1.6~eV (775~nm) in order to directly excite the \textit{A}-exciton transition at room temperature\cite{Frindt1963} at the K (and K$'$) points of the Brillouin zone\citep{Bertoni2016}.

\subsection{\label{sec:Microscope} Time-of-flight momentum microscope}

\begin{figure}
\center{\includegraphics[width=1.0\columnwidth]{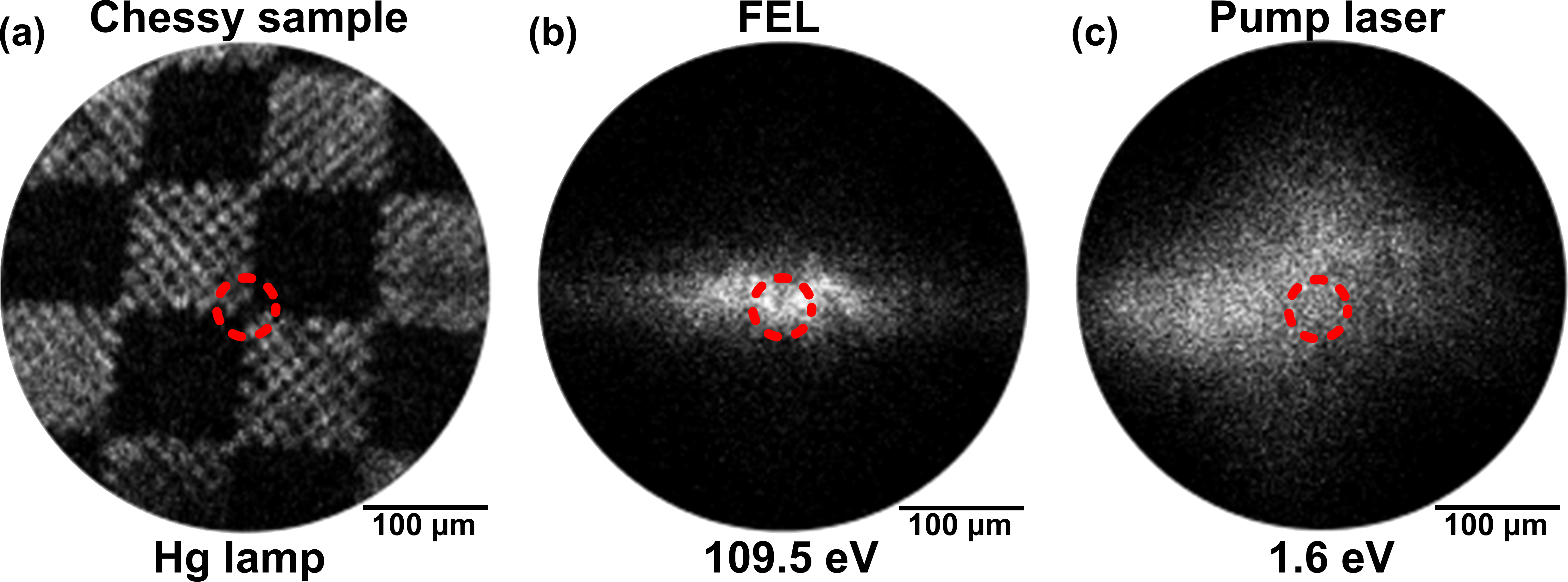}}
\caption{(color online) Spatial overlap determination in the PEEM mode of the momentum microscope using the Chessy test sample. (a) Chessy test sample illuminated by a Hg lamp in order to define the size of the field of view, photon footprints of (b) the FEL beam and (c) the optical pump laser. The red dashed circles mark the regions of interest defined by the field aperture.
 \label{Chessy}
 }
\end{figure}

The schematic cross section of the ToF momentum microscope with the imaging spin filter is shown in Fig.\,\ref{Set-Up}. The lens system, developed at the Max Planck Institute for Microstructure Physics in Halle, Germany, is optimized for maximum k-resolution and a large zoom range of the k-image (0.5 \angstrom$^{-1}$ up to beyond 3~\angstrom$^{-1}$ radius). The combination of the lens system with ToF spectrometry was developed at the University of Mainz. The ToF column, well-shielded against magnetic stray fields, consists of an entrance lens system, a low-energy drift section and a time-resolving image detector. The first Fourier image (achromatic reciprocal image) is formed at the back focal plane of the objective lens. This image is transferred and magnified by two lens groups to the entrance of the field-free low-energy ToF section (800~mm long), where the energy dispersion takes place with typical drift energies of 10--30~eV. The imaging spin-filter crystal could be inserted under 45$^{\circ}$ after two lens groups\cite{Kutnyakhov2016}. Thus, the second field-free low-energy ToF section (vertical branch) is used for energy dispersion, providing spin resolved data (not discussed in this paper) complementary to spin-integral branch data. A set of nine selectable and adjustable field apertures in an intermediate real-space image plane select a well-defined region of interest (ROI) on the sample surface with diameters down to the micrometer range without the need of a fine-focused photon beam. This (real-space) field aperture serves as a contrast aperture for the reciprocal image. For the presented measurements, the field aperture was set to a circular acceptance range of 54~$\mu $m diameter on the sample surface (marked by the red dashed circles in Fig.\,\ref{Chessy}), matching the homogeneous distribution of spatially-overlapped FEL and optical pump laser beams. The energy resolution for operation with the FEL beam was determined to be 130 meV, evaluated from Gaussian broadening of W~4$f$ peaks. The Fermi edge of an FeRh sample revealed a width of 150~meV FWHM \citep{Pressacco2019}, whereas the Fermi edge taken from a Ag(110) sample used as a substrate for a pentacene thin film revealed a width of 78~meV FWHM\citep{Scholz2019}. The increase from the base resolution of the instrument of $<$20~meV (as determined at low energies) can be attributed to contributions from the FEL-induced space-charge effects as well as to electronic noise and time jitter in the ToF measurement. The momentum resolution in the experiment at FLASH was 0.06~\angstrom$^{-1}$, measured for the Cu(100) valence bands. Using similar electron optics, a momentum resolution of 0.005~\angstrom$^{-1}$ was demonstrated by Tusche \textit{et al.} in a lab experiment \citep{Tusche2015}.

The sample was mounted on a hexapod manipulator for precise adjustment of six coordinates, ($\Delta x$, $\Delta y$, $\Delta z$, $\Delta \theta$, $\Delta \phi$, $\Delta \varphi$), with the possibility of He-cooling and high temperature flashing by electron bombardment. All measurements shown here have been performed with the sample kept at room temperature. The single-crystalline 2H-WSe$_2$ sample with an approximate size of $5 \times 5 \times 0.2$~mm$^3$ was purchased from HQ Graphene. The crystal was fixed onto the sample holder using conductive epoxy glue (EPO-TEK H20E) and was prepared \textit{in-situ} by mechanical exfoliation using adhesive tape within the load-lock prior to the measurements and immediately transferred into the measurement chamber with a base pressure better than $2 \times 10^{-10}$~mbar.

\subsection{\label{sec:Recording} Multidimensional recording scheme}

The key element of the detection system is the 3D ($k_x$,$k_y$,$t$)-resolving delay-line detector (DLD) \cite{Oelsner2001,Oelsner2010}. The standard single-channel DLD (temporal resolution 150~ps, 80~mm active area, spatial resolution $\sim$80~$\mu $m) has a limited multi-hit detection capability restricting the effective acquisition rate to typically below one electron per photon pulse. For the conditions of the present experiment, when the number of photoemitted electrons per pulse is close to the space-charge limit and the electron collection efficiency is extremely high, the detector becomes the bottleneck of the time-resolved photoemission experiment. Given the 3D recording scheme of the momentum microscope itself, the detection efficiency can be improved on the detector side by increasing its multi-hit capability\citep{Fognini2014}. For the presented experiment, we have used a segmented 4-quadrant DLD (DLD8080-4Q) with an active area of 80~mm in diameter. The boundaries of each quadrant are clearly visible in the energy-momentum cuts shown in Fig.\,\ref{Ernstorfer}(d). This DLD can ideally detect four counting events arriving at the four quadrants with independent readout channels. For random event distribution with Poisson statistics the average value reduces to 1--2 electrons per pulse. At the same conditions a single-channel DLD can record at maximum 0.5 electrons per pulse.

In time-of-flight instruments electrons with very low kinetic energies cause a background, which can extend over the entire spectrum. In the present microscope we use one of the lens elements as high-pass filter, cutting off kinetic energies below a certain threshold. This cutoff restricts the width of the energy spectrum in the drift tube to values from $\sim$15~eV down to 7~eV (the latter case is discussed quantitatively by Tusche et al. in Ref. \onlinecite{Tusche2016}). At a typical drift energy in the ToF section of 35~eV (for electrons at E$_F$) the resulting range of kinetic energies E$_{kin}$= 20--35~eV is reduced to E$_{kin}$= 28--35~eV. Hence, this high-pass lens element avoids slow electrons entering the drift tube and, thus, they do not contribute to total number of detected electrons per pulse.

The multidimensional recording scheme at high repetition rate SASE-type FELs is very demanding. In the general case, including spin analysis, two vector quantities (momentum \textbf{k} and spin polarization $\boldsymbol{\sigma}$) as well as at least three scalar quantities (binding energy E$_B$, pump-probe delay $\tau_{delay}$, and SASE jitter $\tau_{SASE}$) must be recorded together with the photoemission intensity, I(E$_B$, \textbf{k}, $\boldsymbol{\sigma}$, $\tau_{delay}$, $\tau_{SASE}$). Additionally, in the case of dichroism measurements, recording of the position of the quarter-wave plate and the polarization of the pump laser increases the number of relevant quantities. Since the absolute time of an individual photon pulse in a macrobunch pulse train contributes to the actual pump-probe delay, all these parameters have to be recorded simultaneously for each individual photoelectron event. In the present paper, we do not show spin-resolved data and k$_z$ is not relevant for the quasi-2D system of WSe$_2$. Hence, the number of parameters to be recorded is five, i.e., (k$_x$, k$_y$, E$_B$, $\tau_{delay}$, $\tau_{SASE}$). The momentum coordinates (k$_x$, k$_y$) are directly resolved via the full-field imaging, the binding energy E$_B$ is obtained from ToF analysis, the pump-probe delay $\tau_{delay}$ is set by the delay stage of the pump laser, and $\tau_{SASE}$ as well as other FEL pulse parameters are recorded by the arrival time monitor on a pulse-to-pulse basis. All the data are fed into the FLASH data acquisition (DAQ) system\cite{FLASH-DAQ}.

Additionally, the single-shot detection offered by the DLD allows each photoemitted electron to be correlated to a specific FEL or an optical pulse. This is in particular important for the SASE process of the FEL, where the arrival time and intensity of the X-ray pulses fluctuate from pulse to pulse. Thus, data acquired by the DAQ can be post-processed and corrected, e.g., for intensity fluctuations within the FEL bursts (according to gas monitor detector (GMD) values), or for internal timing jitter of the FEL measured by the beam arrival monitor (BAM) with single-pulse precision\citep{Viti2017}, or for the power of the optical laser for each pumped pulse, respectively.

In order to process the large amount of data acquired with the ToF momentum microscope and analyze them subsequently, we have developed a distributed work-flow, based on efficient interaction with single-event data and compatible with data acquisition schemes at both 3$^{rd}$ and 4$^{th}$ generation light sources as well as table-top laser sources. In addition, we define the corresponding format for data storage and reuse and compare visual representations pertaining to the band mapping data. Our approach provides a basis for the standardization, sharing and integration of band structure mapping data into materials science databases\citep{Xian2019,Zenodo2019}.

\subsection{\label{sec:Overlap} Spatial and temporal overlap}

The real-space imaging mode (photoemission electron microscopy, or PEEM) of the instrument is used to analyze the region of interest, in particular for inhomogeneous or structured samples (Fig.\,\ref{Chessy}(a)). Using threshold photoemission (e.g., by a Hg lamp or UV laser diode) the spatial resolution of the microscope is determined to be $<$50~nm. For metallic samples this mode allows us to avoid spots of intense plasmonic electron emission, induced by the pump laser beam, by moving such regions out of the field of view (FoV) using the hexapod manipulator. The PEEM mode of the momentum microscope is also ideally suited to find the position of the FEL beam and to align the whole setup during the initial phase of the experiment and to control the spatial overlap of pump and probe pulses. This is done by means of the fully automated and adjustable ($X$, $Y$, $Z$, $\Omega$) dedicated support frame, in order to align the whole setup with respect to the FEL beam position in the middle of the FoV of the momentum microscope (Fig.\,\ref{Chessy}(b)). Alignment, focusing and setting up of precise spatial overlap of the pump-laser beam (Fig.\,\ref{Chessy}(c)) (for visualization purposes the pump laser beam is set slightly off center to show the sensitivity and easiness of spatial overlap alignment) is achieved with the help of motorized steering mirrors and lenses in the optical laser beamline and the laser in-coupling system. Spot sizes of both beams are defined by measuring the footprint of secondary electrons emitted from the sample. The length scale is conveniently calibrated by means of a test object like the one shown in Fig.\,\ref{Chessy}(a) (Chessy, PLANO GmbH).

\begin{figure}
\center{\includegraphics[width=1.0\columnwidth]{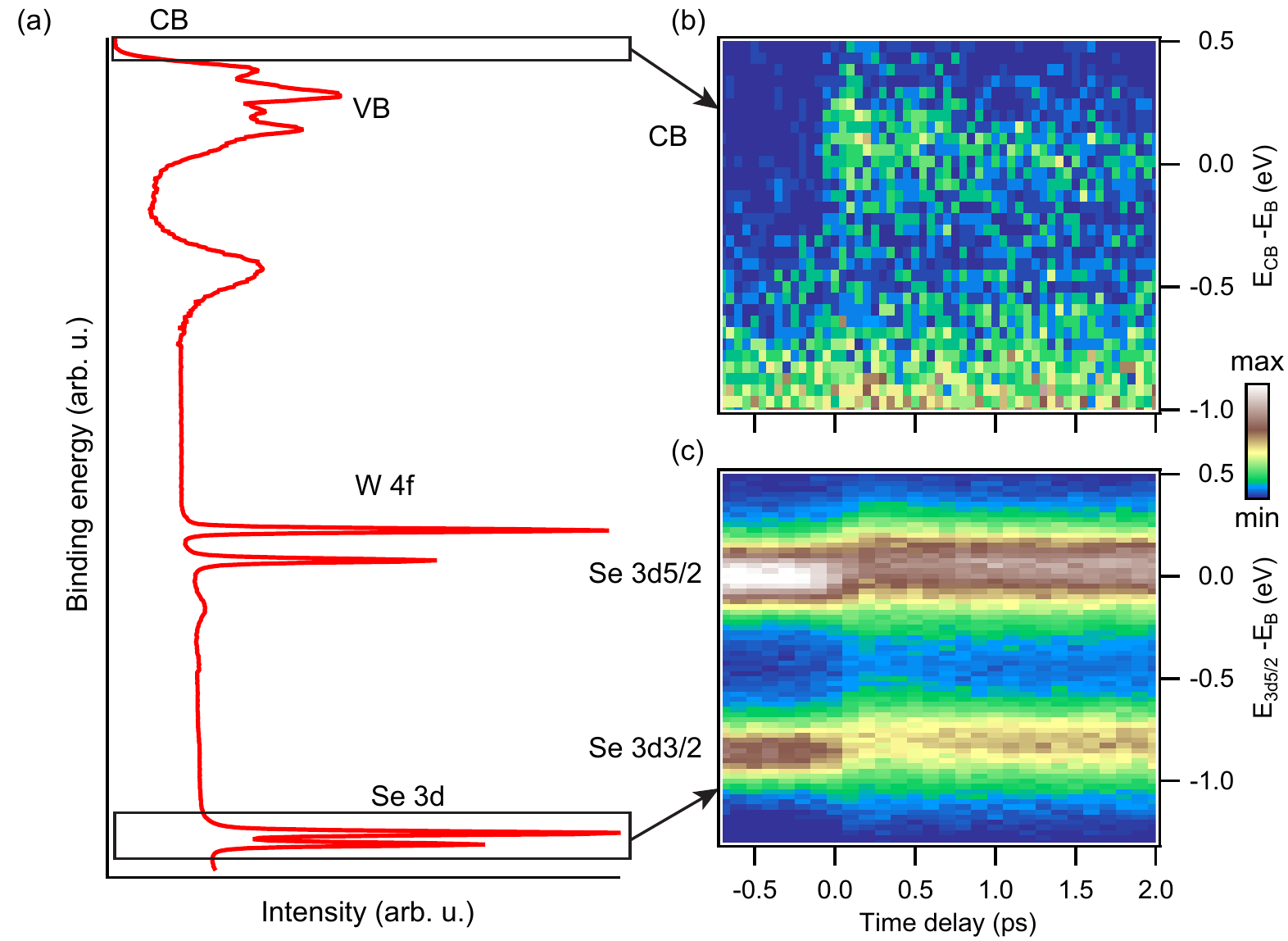}}
\caption{(color online) Time-resolved dynamics in a broad energy range of the photoemission spectrum of WSe$_2$. The W~4$f$ and the Se~3$d$ core levels lie about 33~eV and 55~eV below the Fermi level\cite{Shallenberger2018}, respectively. (a) Momentum-integrated photoemission spectrum recorded by the momentum microscope with the FEL pulses at 109.5~eV (3$^{rd}$ harmonic). Time-resolved measurements are conducted with 775~nm linear $s$-polarized optical pulses as pump, showing concurrent dynamics in (b) the WSe$_2$ conduction band and (c) Se $3d$ core level doublets.
\label{trXPS}
}
\end{figure}

\begin{figure*}
\center{\includegraphics[width=16cm]{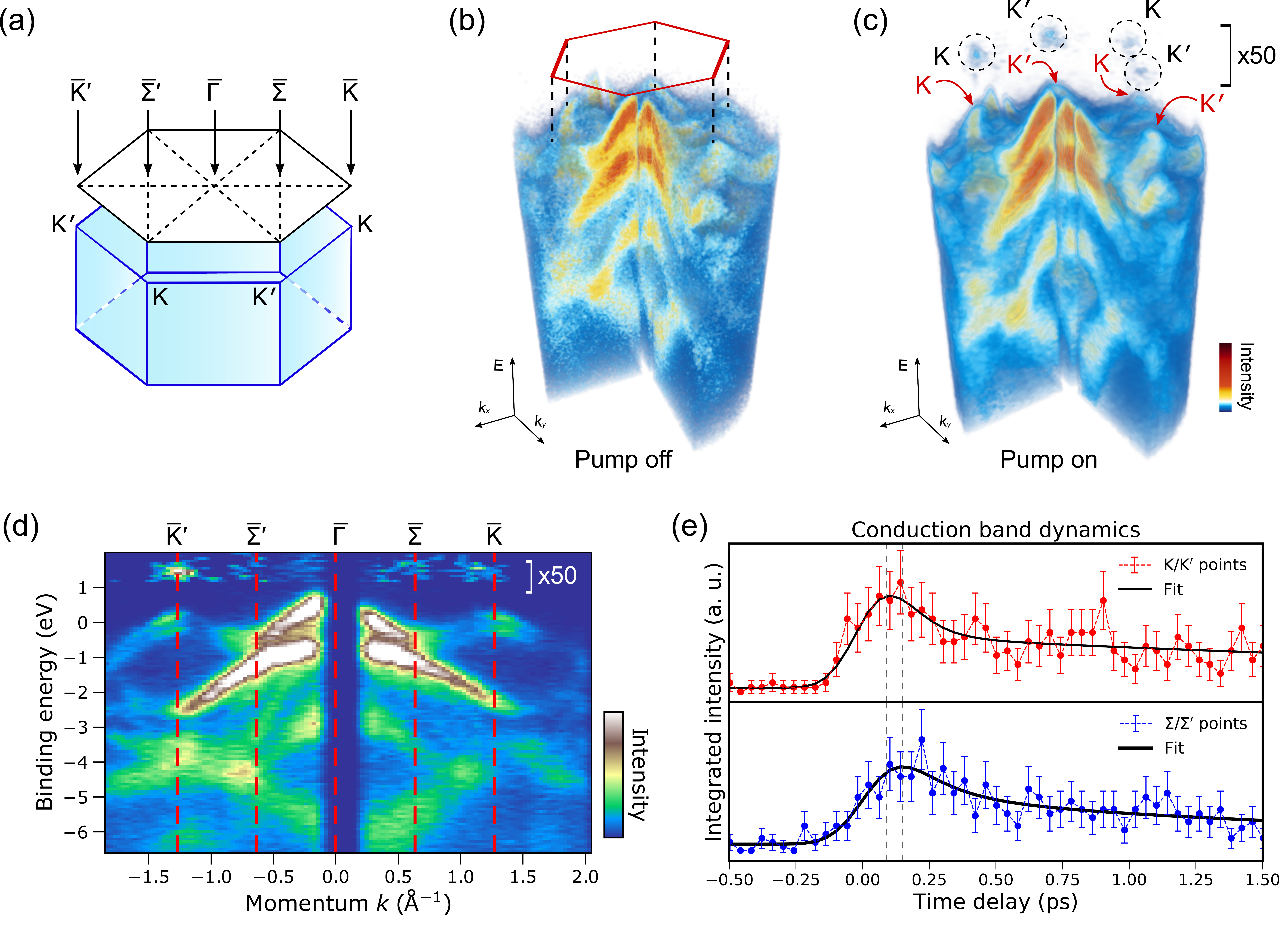}}
\caption{(color online) Photoelectron momentum microscopy of the equilibrium and nonequilibrium electronic band structure of semiconducting WSe$_2$. (a) The bulk (blue) and surface (white) Brillouin zone of WSe$_2$. (b-c) Volumetric rendering of the band-mapping measurement outcome before (b) and after (c) optical excitation with 775~nm pulses. The data from one quadrant is cut out to reveal the dispersive electronic bands on the inside. (d) A momentum path sampled through the volumetric data along the $\overline{\text{K}}'$--$\overline{\Gamma}$--$\overline{\text{K}}$ momentum path labeled in (a). The vertical cut-out around 0 \angstrom$^{-1}$ in (d) removes the influence from the edges of the quadrant detector. (e) Temporal evolution of the excited state signal integrated over a region around the K/K$^{'}$ and $\Sigma$/$\Sigma^{'}$ points in the first conduction band. The fittings use a model function of a double-exponential decay convoluted with the system response function. The signals in the conduction band K/K$^{'}$ and $\Sigma$/$\Sigma^{'}$ valleys reach their respective maxima with a delay of $\sim$60 fs.
\label{Ernstorfer}
}
\end{figure*}

Once the spatial overlap is set, the temporal overlap between pump and probe pulses, i.e., the value for “time zero” $t_0$, has to be found. For a coarse overlap determination, we use an in-vacuum antenna (inner conductor of a large-bandwidth cable, hit by both beams) in combination with a fast oscilloscope, which can pin down the overlap in the sub-50~ps range. To determine the temporal overlap with sub-ps precision, we operate the momentum microscope in the k-imaging mode and set the sample potential such that the detectable energy range is located close to the Fermi edge or a core level (depending on the sample material characteristics and the excitation energy). Then, by scanning the delay stage position with an appropriate fluence of the pump laser beam, one observes excited state dynamics above the Fermi edge (see Fig.\,\ref{trXPS}(b) for momentum-integrated and Fig.\,\ref{Ernstorfer}(c),(d) for momentum-resolved data) in the case of trMM on WSe$_2$ or pump-induced changes in core levels (see Fig.\,\ref{trXPS}(c) for Se $3d$ core level dynamics around $t_0$) in case of trXPS. Integrating over the whole Brillouin zone (BZ) in the reciprocal space helps to speed up the quite challenging and time-consuming process of $t_0$ finding. Particularly, we are not limited to an excited signal from a single high-symmetry point (e.g., $\Gamma$- or K-point) or from certain directions in reciprocal space (e.g., K$'$-$\Gamma$-K plane as shown in Fig.\,\ref{Ernstorfer}(d)).

\subsection{\label{sec:SpaceCharge} Space-charge effect}

One of the major obstacles in trPES experiments is the Coulomb repulsion of the electrons in the beam, leading to the space-charge effect. It can dramatically affect the energy and momentum resolution that can be reached in a pump-probe photoemission experiment. The space-charge effect has previously been studied using conventional dispersive spectrometers at FLASH\citep{Hellmann2012,Hellmann2012a} and the FEL SACLA in Japan\citep{Oloff2014}. In this section, we discuss the characteristics of the space-charge interaction in a momentum microscope, where the strong extractor field in front of the sample causes a behavior fundamentally different from the conditions in conventional spectroscopy, where a field-free region exists close to the sample\citep{Rossnagel2018}.

In addition to the deterministic interaction of a given electron with the average charge distribution of all other electrons (usually termed the space-charge effect), there are individual electron-electron scattering events, which are stochastic\citep{Schoenhense2015a}. In order to understand the details of this effect in a momentum microscope, it is necessary to distinguish the different species of electrons contributing to the total photoyield. The electrons of interest are the photoelectrons emitted from the valence band or from a core level by the FEL pulse. This signal of (fast) electrons is accompanied by the secondary electrons, which can outnumber the true photoelectrons by orders of magnitude, depending on the material and the energy of the fast photoelectrons and possible further core level electrons, whose binding energies are lower than the photon energy. The Coulomb repulsion induced by the slow electrons causes a non-uniform acceleration and gives rise to an increase of kinetic energy for the electrons of interest. The increase in kinetic energy is maximum on the optical axis and decreases with increasing distance from the axis, i.e., with increasing parallel momentum. The spatial profile can be approximated by a two-dimensional Lorentzian function. This part of the space-charge effect is deterministic and can be corrected numerically\citep{Schoenhense2018a}. In a conventional spectrometer the repulsion is uniform and hence the Lorentzian deformation is absent. 

The pump-laser pulses can release a third species of electrons via multiphoton photoemission (nPPE). Their intensity rises nonlinearly with the n-th power of the pump fluence until  saturation effects become significant. At very high fluences the high charge density in front of the surface can lead to a Coulomb blockade effect. Due to the difference in kinetic energies between the photoelectrons and the pump-induced low energy electrons, the Coulomb forces exerted on the photoelectrons by the pump-induced electrons strongly depend on the pump-probe time delay\citep{Ulstrup2015,Al-Obaidi2015}. For negative time delays, i.e., when the pump pulse excites the sample after the probe pulse, the space-charge induced energy shift will always be accelerating, because the slow electrons trail behind the fast photoelectrons. However, for positive time delays (as relevant in pump-probe experiments), i.e., when the pump pulse hits the sample before the probe pulse, the photoelectrons will pierce through the cloud of slow pump-induced electrons at some instant in time. Therefore, the space-charge effect will, at first, slow down, and later accelerate the photoelectrons, leading to positive and negative energy shifts, depending on the pump-probe delay. In principle, as long as we neglect individual scattering events this behavior is still deterministic and can be modeled by simulations.

\begin{figure}
\center{\includegraphics[width=1.0\columnwidth]{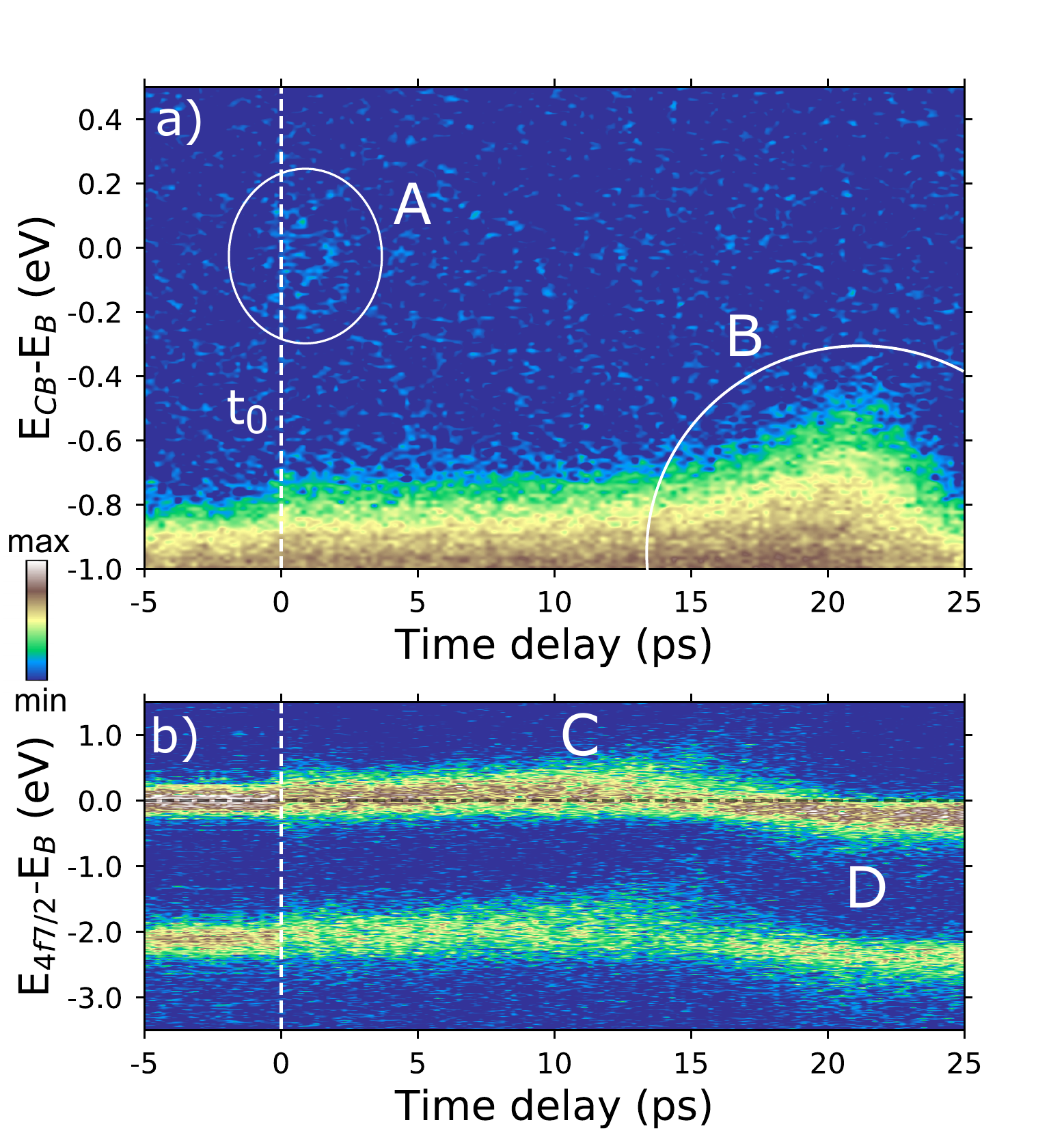}}
\caption{(color online) Space-charge-induced features measured by the momentum microscope. (a) Momentum-integrated intensity of WSe$_2$ conduction band dynamics measured with 775~nm linear $s$-polarized optical pulses as pump and the FEL pulses at 36.5~eV (fundamental) as the probe. The true time-zero energy shift is visible at temporal coincidence at 0~ps in the excited state signal marked by A (white circle). The large shift at 21~ps marked by B (“long-range space-charge” shift) originates from the space-charge interaction in the electron-optical column. (b) W $4f$ core level dynamics measured with 775~nm linear $s$-polarized optical pulses as a pump and the FEL pulses at 109.5~eV (3$^{rd}$ harmonic) as the probe. Characteristic long-range space-charge bipolar features marked with C and D.
 \label{SpaceCharge}
 }
\end{figure}

However, even at moderate intensities, stochastic electron-electron scattering happens, in particular, when the fast photoelectrons penetrate the dense cloud of slow electrons. This type of interaction leads to an energy broadening and to a randomization of the k-distribution, i.e., to a loss of momentum information. The number of slow electrons excited by the pump laser (photon energy 1.5--3~eV) is material dependent and for a given material strongly depends on the surface quality. In particular, surface inhomogeneitites on metal surfaces can act as hot spots of strong plasmonic electron emission. Unlike the deterministic forces, it is generally impossible to correct in postprocessing the effect of stochastic scattering. The only compensation is to reduce the signal of undesired slow electrons. Plasmonic hot spots can easily be driven out of the FoV by lateral sample shift, their position can be monitored in the PEEM mode (cf. Fig.\,\ref{Chessy}). A large work function and small electron density of states near the Fermi level are favorable to minimize the amount of photoexcited slow electrons. 

The long-range space-charge shift can be exploited for a coarse determination of the temporal overlap. The extractor field and the lens optics confine the cloud of slow electrons in a small spatiotemporal segment of phase space. When propagating through the electron-optical column, this macrocharge of slow electrons exerts long-range forces on the fast photoelectrons over macroscopic distances of more than 50~mm as concluded from recent simulations\citep{Schoenhense2018a}. The net action on the photoelectrons can be accelerating or decelerating, depending on the time delay between the pump and the probe pulses. This additional energy shift occurs at larger positive time delays and, due to the pronounced characteristic bipolar feature, this effect could be easily mistaken for the dynamics at "true" $t_0$. However, since the bipolar energy shift extends over a much larger range of pump-probe time delays, it can be distinguished from the "true" ultrafast dynamics at around the temporal overlap between the pump and the probe pulses and, therefore, serve as a reference point in the time delay range for systems with a small excited-state signal at $t_0$. At the settings we used in the experiment, the time differences between "time zero" and the long-range space-charge-induced energy shift amounts to about 20~ps. 

A typical temporal behavior of space-charge-induced shift for the total k-integral signal is shown in Fig.\,\ref{SpaceCharge}. For the valence band of WSe$_2$ in Fig.\,\ref{SpaceCharge}(a), we see both the time-zero shift at $t_0$ with its characteristic signal A (see next section) and the long-range shift culminating around 21~ps, labeled as signal B. At the setting used for the W $4f$ core level doublet in Fig.\,\ref{SpaceCharge}(b), the positive shift is maximized at 14~ps (signal C), reverses its sign at 16~ps and reaches its negative extremum at 22~ps (signal D). In both cases, $t_0$ is far away and nicely separated from the long-range space-charge shift. The result in  Fig.\,\ref{SpaceCharge}(b) is typical, but the shift depends on many intrinsic (microscope setting, e.g. the extractor voltage, i.e. the potential between the sample and the first electrode of the momentum microscope) and extrinsic parameters (sample, presence of hot spots, excitation energy of the probe beam) and can be moved on the pump-probe time delay axis accordingly to changing any of them. For a given sample the extractor setting has a clear "deterministic" influence, which allows to identify the in-column space charge shifts as discussed in Ref. \onlinecite{Schoenhense2018a}.

We summarize this section as follows: The stochastic part of the Coulomb interaction between the different species of electrons sets a principal limit to all fs pump-probe photoemission experiments (see also Ref.\,\onlinecite{Oloff2014}). The deterministic part, however, can be exploited as a (very fast) mechanism to determine the temporal overlap. This nicely complements the fast adjustment of the spatial overlap, the elimination of surface defects via lateral shift and the definition of the region of interest by the field aperture using the PEEM mode of the instrument. 

\section{\label{sec:Results}Results and discussion}

In momentum space, {WSe}$_2$ exhibits a hexagonal BZ with alternating K and K$'$ (or K$^+$ and K$^-$) points as the hexagonal vertices, see Fig.\,\ref{Ernstorfer}(a). The electronic spin characters in the vicinity of the K and K$'$ points are locked to their corresponding valence and conduction band extrema, i.e., valleys, a phenomenon that exists in monolayers as well as in individual layers of the crystalline bulk \citep{Mak2012,Zeng2012,Wang2018,Zhang2014,Bertoni2016}. Time- and momentum-resolved photoemission is a unique tool for the investigation of layer-resolved electron dynamics due to its surface sensitivity. In our series of experiments, we optically excite the samples with 775~nm light pulses (pump) and probe by mapping the electronic band structure via photoemission with 36.5~eV soft X-ray pulses delivered at the FLASH PG2 beamline \cite{Martins2006,Wellhoefer2007,Gerasimova2011}. The pump pulses are resonant with the \textit{A}-exciton transition\cite{Frindt1963} in {WSe}$_2$ and populate the conduction band K and K$'$ valleys, where we follow the subsequent dynamics. 

We used linearly as well as circularly polarized 775~nm pump pulses to measure the transient electronic structure of the valence band and the excited states resolved in energy and both parallel momentum directions (see Fig.\,\ref{Ernstorfer}(b)-(c)). Excitation of the \textit{A} exciton generates excited state population initially localized in the K and K$'$ valleys, which rapidly scatters to the energetically lower $\Sigma$ and  $\Sigma'$ valleys, respectively. Our results are in good agreement with previous measurements using a hemispherical detector that samples selected $k$-$E$ slices in the momentum space \citep{Bertoni2016}, but now, the direct and simultaneous recording in 3D parameter space ($k_x$, $k_y$, $E$) yields a more complete characterization of the excited-state electronic distribution, while requiring only a fraction of the measurement time compared with conventional spectrometers. Therefore, it gives us the possibility to access different high symmetry points in parallel without any sample manipulation. 
Thus, we are able to observe and compare the conduction band carrier distribution in the non-equivalent K (or K$^{'}$) and  $\Sigma$ (or $\Sigma{'}$) points. The temporal evolution of the photoemission intensity in the conduction band, shown in Fig.\,\ref{Ernstorfer}(b)-(c), is plotted in Fig.\,\ref{Ernstorfer}(e) for the experiment with linear $s$-polarized pump pulses. 
A numerical fit of the population dynamics at K/K$^{'}$ points yields a fast decay component with a time constant of $\sim$80~fs. The signal at the $\Sigma$/$\Sigma^{'}$ points reaches its maximum with a  delay of $\sim$60 fs compared to the K valley signal and shows a multiexponential relaxation dynamics with a fast time constant of $\sim$160~fs. The delay is characteristic of the temporal ordering of intervalley scattering in this class of materials\cite{Bertoni2016}. The fittings use a model function of a double-exponential decay convoluted with the system response function. The system response function, i.e.~the effective pump-probe cross-correlation, was determined by fitting the time dependence of the laser-assisted photoemission (LAPE) signal between valence and conduction bands with a Gaussian, resulting in $\sim$150~fs FWHM. We point out that the system response function is shorter than the duration of the soft X-ray pulse as the photoelectrons are only collected from a selected area by means of field aperture within the soft X-ray footprint on the sample (see Fig.\,\ref{Chessy}), mitigating the wavefront-tilt broadening of the probe pulses.
In principle, the approach is capable of detecting dynamics of holes. It requires, however, a very careful analysis, as both the hole formation as well as changes in the spectral function of the remaining occupied states can cause a reduction in the photoemission intensity of the valence band states. The separation of these two contributions is not well established yet and will be addressed in future work.

The benefit of FEL radiation is the additional access to core levels\citep{Dendzik2019}, giving information on transient charge redistributions and XPD patterns\citep{Curcio2019}, adding time-resolved structural information. The straightforward access to the interplay of the ultrafast structural and electronic responses is a stronghold of our technique.
To demonstrate this potential of the spectral tunability of the FEL, i.e.,~the availability of higher harmonics in addition to the fundamental output and the capability of the monochromatized PG2 beamline to easily switch between them, trXPS as well as trXPD measurements have been performed using the same setup and the same {WSe}$_2$ sample as for trMM studies. As shown in Fig.\,\ref{trXPS}, femtosecond dynamics of the Se $3d$ core level photoemission signal were measured with 775~nm optical pump pulses and the FEL probe pulses at 109.5~eV (3$^{rd}$ harmonic) photon energy. We observe that pumping of the \textit{A}-exciton resonance induces pronounced changes of the Se $3d$ core level position and lineshape, see Fig.\,\ref{trXPS}(c). More details of the trXPS results are discussed by Dendzik \textit{et al.}~in Ref. [\onlinecite{Dendzik2019}]. Thus, the dynamics of the valence electrons (trMM) could be compared with the concurrent changes induced in the core level spectral function simply by combining two experiments in one. This is performed by only changing:
\begin{itemize}[noitemsep]
\item sample bias -- to shift the focus of the momentum microscope from the valence band region to core levels;
\item monochromator settings -- to go from the fundamental to the third harmonic radiation;
\item gas attenuator or solid state filters -- to be able to attenuate either the fundamental or the third harmonic radiation.
\end{itemize}
Moreover, by setting the momentum microscope to the maximum possible k-space FoV (up to 7~\angstrom$^{-1}$ in diameter) and measuring the core level spectra with momentum resolution, it is possible to perform trXPD measurements. Thus, the experimental setup described here can additionally provide information on the structural dynamics of the WSe$_2$ sample induced by the optical excitation.
Another point is to exploit the k$_z$-dependence or 4D-recording of the full bulk band structure via photon-energy sequences\citep{Medjanik2017}. This approach requires a continuously-tunable photon-energy range of about 200~eV, as available at FLASH and other FEL sources. These advantages contrast with the low repetition rate. This efficiency-limiting parameter will increase to up to 27.000 pulses per second at the European XFEL and other forthcoming continuous-wave FEL sources. The specific advantages of the instrument will be leveraged by the current technical developments of FEL sources worldwide.

Nevertheless, the necessary accumulation time also strongly depends on the sample. In the present paper the k-integral measurement shown in Fig.\,\ref{SpaceCharge} was taken in 10 minutes (for FeRh\citep{Pressacco2019} a similar pattern took just one minute). The measurement run shown in Fig.\,\ref{trXPS} was taken in 70 minutes, and the full 3 D data set of Fig.\,\ref{Ernstorfer} required 5 hours. A full 4D time-resolved photoelectron-diffraction data set\citep{Curcio2019} was acquired in 40 hours.

\section{\label{sec:Summary}Summary and conclusion}

We present a novel setup combining full-field momentum imaging and time-of-flight energy recording (the 3D data recording architecture) with a sample preparation chamber and many characterization tools. The instrument is the ideal tool for very time-consuming pump-probe photoemission experiments with the fs radiation from the high repetition rate free-electron laser FLASH at DESY, Hamburg. The high FEL photon energies of up to 830~eV pave the way for time-resolved experiments in core level spectroscopy, photoelectron diffraction as well as valence band spectroscopy with varying probing depth (via photon energy tuning). In this respect, the experimental opportunities enabled by the instrument described here go far beyond the capabilities of table-top HHG sources. With the present setup, it is possible to investigate ultrafast electronic dynamics in a wide range of chemical and physical systems by means of high resolution time- and momentum-resolved photoemission spectroscopy.

Overall, the simultaneous detection of the electronic, spin, and geometric structure in the time domain, with femtosecond resolution, will open a new avenue for the direct determination of the couplings between electronic, spin, and lattice degrees of freedom that give rise to the fascinating emergent properties of quantum materials. The parallel image acquisition further paves the way to single-shot experiments at ultra-bright fs-sources. Employing time- and momentum-resolved photoemission spectroscopy on the layered semiconductor WSe$_2$, we observed spin-, valley-, and layer-polarized excited state populations upon excitation to K/K$^{'}$ valleys with linearly and circularly polarized pump pulses, followed by ultrafast ($<$ 100~fs) scattering of carriers towards the global minimum of the conduction band in the $\Sigma$/$\Sigma^{'}$ valleys. First preliminary results with an imaging spin-filter as described in Ref. [\onlinecite{Kutnyakhov2016}] suggest that femtosecond pump-probe spin mapping is feasible at the given conditions.

Different detector development steps were performed in order to improve the multi-hit capability of the DLD. The 4Q-DLD detector, discussed above, was upgraded to an improved version with 8 segments (DLD6060-8s). It comprises a stack of two 4-quadrant DLDs rotated by 45$^{\circ}$ with respect to each other and provides an active area of 60 mm. The new device can detect 2.5--3 electrons per FEL pulse with much improved resolution at the boundaries between all segments. Data using this detector will be published in forthcoming papers\citep{Scholz2019,Pressacco2019,Curcio2019}. Moreover, new delay-line architectures with massive parallelization (up to 256 individual delay lines with fully independent readout channels) are currently under commissioning at Surface Concept GmbH\citep{Turcato2014}.

\begin{acknowledgments}
This paper is dedicated to our dear colleague Wilfried Wurth who passed away unexpectedly on May 8, 2019. Wilfried Wurth initiated and founded the time- and momentum-resolved photoemission (HEXTOF) project at FLASH as an experimental base of this work, he pushed it and took great care of it. We appreciate and miss now the uncounted high quality and inspirational scientific and technical discussions with Wilfried, especially during the beam times. We miss his outstanding input to our work and his tremendous impact and dedication to all the progress and achievements made at FLASH in general.

We acknowledge the support of scientific and technical staff of DESY (Hamburg, Germany), a member of the Helmholtz Association HGF, operating FLASH in an excellent way, in particular the machine operators and run coordinators taking care of optimum FEL conditions and for making the experiment possible. The work was funded by the DFG and performed within the framework of the SFB 925 (project B2) as well as the Transregio SFB/TRR 173 Spin+X, RE1469/13-1 and the TRR 227 Ultrafast Spin Dynamics (projects A09 and B07). This project has received funding from the BMBF (05K16UM1, 05K13GU3, 05K16PGB), the European Research Council (ERC) under the European Union's Horizon 2020 research and innovation program (Grant Agreements No. ERC-2015-CoG-682843), and the Max Planck Society. F.P.~acknowledges funding from the excellence cluster “The Hamburg Centre for Ultrafast Imaging - Structure, Dynamics and Control of Matter at the Atomic Scale” of the Deutsche Forschungsgemeinschaft (DFG EXC 1074). L.R.~acknowledges funding from the DFG in the Emmy Noether program under grant number RE 3977/1. P.H., J. M. and S.U acknowledge funding from the Danish Council for Independent Research, Natural Sciences under the Sapere Aude program (Grants No. DFF-4002-00029 and DFF-6108-00409), by VILLUM FONDEN via the Centre of Excellence for Dirac Materials (Grant No. 11744) and via the Young Investigator Program (Grant No. 15375), and by the Aarhus University Research Foundation.
\end{acknowledgments}




\section{\label{sec:References}References}
\bibliography{Kutnyakhov_HEXTOF}

\end{document}